\documentclass[pre,twocolumn,showpacs]{revtex4}
\usepackage{graphicx}
\usepackage{amsmath}
\usepackage{ifthen}
\usepackage{amssymb}
\usepackage{color}
\newcommand{\slaninacolor}{true}
\def\slaninafigdir{.}
\ifthenelse{\equal{\slaninacolor}{true}}{%
\def\slaninacolorinname{-color}
\def\slaninacolorinnameunder{_color}

}{
\renewcommand{\color}[1]{{}}

\def\slaninacolorinname{}
\def\slaninacolorinnameunder{}
}
\begin{document}
\title{%
Inelastically scattering particles and wealth distribution in an open economy
}%
\author{Franti\v{s}ek Slanina}
\affiliation{Institute of Physics,
 Academy of Sciences of the Czech Republic,\\
 Na~Slovance~2, CZ-18221~Praha,
Czech Republic%
}
\email{slanina@fzu.cz}
%
\begin{abstract}
Using the analogy with inelastic granular gasses we introduce 
a model for  wealth exchange in society.
The dynamics is governed by a kinetic equation, which allows for 
self-similar solutions. The scaling function has a power-law tail, the
exponent being given by a transcendental equation. In the limit of 
continuous trading, closed form of the wealth distribution is calculated 
analytically.\end{abstract}
\maketitle
%
%
%

%
\section{Introduction}

The distribution of wealth among individuals within a society was one
of the first ``natural laws'' of economics
\cite{pareto_1897}. Indeed, its study was motivated by the desire to
bring the accuracy attributed to natural sciences, namely physics, to
economic sciences. The celebrated Pareto law states that the higher
end of the wealth distribution follows a power-law $P(W)\sim
W^{-1-\alpha}$ with exponent $\alpha$ robust in time.

The validity of the Pareto law was questioned and re-examined many
times but the core message, stating that the tail of the distribution
is a power law remains in force. There are recent investigations,
e. g. 
\cite{lev_sol_97,dra_yak_01a,ree_hug_02,aoy_sou_fuj_03}, giving
reasonable empirical evidence for 
it. In fact, it is not so much the functional form itself but its
spatial and temporal stability that is intriguing. 
Indeed, while the value of the
exponent $\alpha$ may slightly vary from one society to another, the
very fact of the power-law 0.tail in the distribution is valid almost
everywhere. Recent investigations suggest that the range of validity
of the Pareto law may extend as far in the past as to the ancient
Egypt of the Pharaohs 
\cite{abulmagd_02}.

The universality of the power-law tail is surely a phenomenon asking
for explanation. Recently, there was a lot of effort establishing
finally the multiplicative random processes repelled from zero 
as a mathematical source
of the power-law distributions
\cite{lev_sol_96,lev_sol_96a,bi_ma_le_so_98,solomon_98,solomon_99,hua_sol_00,sol_ric_01b,bla_sol_00,sol_lev_00,hua_sol_01,so_co_97,sornette_97a,sornette_98c,ta_sa_ta_97}.
Alternatively, the killed multiplicative processes as sources of
power-laws were studied in \cite{ree_hug_02}.
However, there are plenty of possible ways how the multiplicative
random processes of this type come onto scene. One of the most studied
implementations were the generalized Lotka-Volterra equations
\cite{solomon_98,solomon_99,hua_sol_00,sol_ric_01b} and the analogy
with directed polymers in random media
\cite{ma_mas_zha_98,bou_mez_00,bur_joh_jur_kam_nov_pa_zah_01}. Both
of these schemes are formalized by a kinetic equation describing the
exchange of wealth between agents and global redistribution of wealth 
which plays the role of repelling from zero. Related approaches were
subsequently pursued by a number of studies 
and simulations 
\cite{souma_00,aoy_nag_oka_sou_tak_tak_00,souma_02,sou_fuj_aoy_01,fuj_sou_aoy_kai_aok_02,cha_cha_00,cha_cha_03,cha_cha_man-all,das_yar_03,pia_igl_abr_veg_01,igl_gon_pia_veg_abr_03,pia_igl_03,igl_gon_abr_veg_03,ana_ish_suz_tom_03,miz_kat_tak_tak_03,miz_tak_tak_03,fuj_dig_aoy_gal_sou_03}.

More recently, empirical studies of the lower end of the wealth axis
showed that the distribution of wealth is rather exponential than
power-law, while the high-wealth tail still remains power-law
\cite{dra_yak_01a,dra_yak-all,yakovenko_03}.
This finding was interpreted as a result of a conservation law for
total wealth, leading to the robust Boltzmann-like exponential distribution,
whatever the random wealth exchange be, in full analogy with the
energy distribution in a gas of elastically scattering molecules.

This, together with older studies within the same spirit
\cite{is_kra_re_98}, lead to the view of economic activity as a
scattering process of agents, analogous to inelastically scattering
particles
\cite{cha_cha_00,cha_cha_03,cha_cha_man-all,sca_pic_wes_02,sca_wes_03,gli_ign_02,sinha_03}.
Indeed, the inelasticity is 
indispensable to explain the 
power-law tail and it is also reasonable to suppose that the total
wealth increases on average. 

The numerical simulations performed to date confirm the emergence of
power-law tail in agent-scattering processes with great
reliability. However, analytic insight is lacking in most of the
studies available today. The main concern of our work is to fill this
gap, providing analytical results at least for a simplified model of
wealth exchange. To comply with the task we will be guided by existing
analytical approaches for models of inelastically scattering particles.

Inelastic scattering of particles was studied thoroughly in the context 
of granular materials \cite{ja_na_be_96a}. The simplest one of the
models used is the Maxwell model, whose inelastic variant was
investigated in detail
\cite{bob_car_gam_00,bob_cer-all,bal_mar_pug_01a,isp_kra_00,benn_kra_00,kra_benna_01,benn_kra_02a,bena_benn_lin_ros_03,ern_bri_02b,ern_bri-everything,benn_kra_03,ant_dro_lip_02,barkai_03}.
More realistic models of granular gasses were also introduced
\cite{bal_mar_pug_01,bal_mar_pug_vul_01} but their full account goes beyond
the topic of this work. The most important conclusion of these studies
is that a self-similar solution of the kinetic equations exist, which
is not stationary in time, but assumes time-independent form after
proper rescaling of the energy. The tail of the scaling function
becomes power-law under certain condition.

The formalism developed for granular gases can be readily adapted for
binary wealth exchange of agents. Indeed, within the  mean-field
version of the Maxwell model the particles scatter randomly one with
another  irrespectively of their positions. This corresponds to
randomly picking pairs of agents for interaction, with no care of the
(possibly complex) structure of their relationships. In reality the
economic activity goes along links in a complex social network
\cite{wa_stro_98,bar_alb_99}. Indeed, recently there were
investigations of the role of network topology in wealth
distribution \cite{igl_gon_pia_veg_abr_03,dim_ast_hyd_03}. 
We may consider the present model as an
approximation of that network by a complete graph.

The main difference from the mean-field Maxwell model is that the
energy of the granular gas decreases by dissipation, while the average
total wealth of the agents increases due to the economic activity. 
The sign of the non-conservation is therefore opposite in the two
cases. While the form of the equations may remain the same, the
solution cannot be directly continued from one domain to another.
Therefore, while the case of dissipation is relatively well
understood, new approaches are needed in the case of production. That
is the aim of the present work.

\section{Interacting agents as scattering particles}

\subsection{Description of the process}

Imagine a society of $N$ agents, each of which possess  certain wealth
$v_i$, $i=1,2,...,N$. Time-to time the agents interact in essentially
instantaneous ``collision'' events, when certain fraction of the
wealth 
can be exchanged. Moreover, we suppose the system is open and the interaction
can catalyze an increase of the total wealth of the two interacting
agents. Indeed, the source of the human wealth lies beyond
our society and the ultimate cause is the energy poured to the Earth
from the Sun. Nonetheless, the external energy is utilized only
through a human activity and we simplify the problem by assuming that
the net increase of wealth happens at the very moments of agents' interaction. 

We also assume that only pairwise interaction occurs. This may be a very
crude assumption, as corporate decisions affect many agents
simultaneously. However, we expect the presence of multilateral
interactions does 
not affect the essential mechanisms in work here.

\begin{figure}[ht]
\includegraphics[bb=0 0 331 388,scale=0.35]{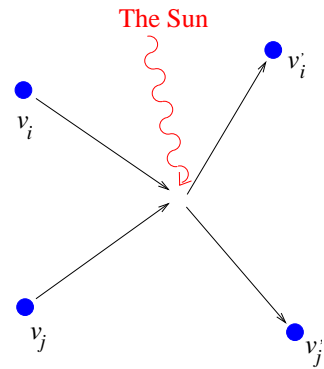}
\caption{%
Schematic picture of the scattering process, where the wealth is
exchanged and produced.  
}
\label{fig:scattering}
\end{figure}

The dynamics of our model is described as follows. In each time step
$t$ a pair
of agents $(i,j)$ is chosen randomly. They interact and exchange
wealth according to the symmetric rule
\begin{equation}
\left(\begin{array}{l} 
v_i(t+1)\\
v_j(t+1)
\end{array}\right)
=
\left(\begin{array}{cc}
1+\epsilon-\beta&\beta\\
\beta&1+\epsilon-\beta
\end{array}\right)
\left(\begin{array}{l}
v_i(t)\\
v_j(t)
\end{array}\right)
\label{eq:wealthexchange}
\end{equation}
All other agents leave their wealth unchanged,
$v_k(t+1)=v_k(t)$ for all $k$ different from both $i$ and $j$. 
The parameter $\beta\in(0,1)$ quantifies the wealth exchanged, while
$\epsilon>0$ measures the flow of wealth from the outside. The process
is sketched schematically in Fig. \ref{fig:scattering}.

This rule is similar to those studied in
\cite{is_kra_re_98,benn_kra_00,bena_benn_lin_ros_03} and simulated
numerically in  
\cite{cha_cha_00,cha_cha_man-all,sca_pic_wes_02,sinha_03} 
but we consider it slightly more realistic as it treats the agents
in a priori symmetric manner. It also embraces various sources of
wealth non-conservation within a single effective parameter
$\epsilon$. In fact, also the formulation based on the similarity with
the problem of directed polymers
\cite{ma_mas_zha_98,bou_mez_00} can
be reduced to a rule of the form similar to
(\ref{eq:wealthexchange}). Therefore, 
we are studying a representative of a whole class of related models
and we expect the analytical results we will present have 
rather broad relevance.

\subsection{Kinetic equation}

The equation (\ref{eq:wealthexchange}) describes a matrix
multiplicative stochastic process of vector variable $v(t)$ in
discrete time $t$. Processes of this type are thoroughly studied
e. g. in the context of granular gasses. Indeed, if the variables
$v_i$ are interpreted as energies corresponding to $i$-th granular 
particle, we can map the process to the mean-field limit of the 
Maxwell model of inelastic particles. However, the energy dissipation 
conventionally quantified by the restitution coefficient implies now
the negative value $\epsilon<0$, contrary to our assumption
$\epsilon>0$. We will see later that this apparently small variation
makes big difference in the analytical treatment of the process.

The full information about the process in time $t$ is contained in the
$N$-particle joint probability distribution $P_N(t;v_1,v_2,...,v_N)$. 
However, we can
write a kinetic equation involving only one- and two-particle
distribution functions
\begin{equation}
\begin{split}
&P_1(t+1;v)-P_1(t;v)
=\frac{2}{N}\bigg[-P_1(t;v)+\\
&+\int P_2(t;v_i,v_j)\,\delta((1-\beta+\epsilon)v_i+\beta v_j-v)\,
\mathrm{d}v_i\mathrm{d}v_j\bigg]
\label{eq:kinetic}
\end{split}
\end{equation}
which may be continued to give eventually an 
infinite hierarchy of
equations of BBGKY type. As a standard approximation we use the 
factorization 
\begin{equation}
P_2(t;v_i,v_j)=P_1(t;v_i)P_1(t;v_j)
\label{eq:molecularchaos}
\end{equation}
which breaks
the hierarchy on the lowest level, neglecting the correlations between
the wealth of the agents, induced by the scattering. In fact, this
approximation becomes exact for $N\to \infty$. Therefore, in
thermodynamic limit the one-particle distribution function bears all
information.

Rescaling the time as $\tau=2t/N$ in the thermodynamic limit
$N\to\infty$, we obtain for the one-particle distribution function
$P(\tau;v)=P_1(t,v)$ a 
Boltzmann-like kinetic equation
\begin{equation}
\begin{split}
&\frac{\partial P(v)}{\partial \tau}+P(v)
=\\
&\;\int P(v_i)P(v_j)\,\delta((1-\beta+\epsilon)v_i+\beta v_j-v)\,{\rm
d}v_i{\rm d}v_j
\label{eq:boltzmann}
\end{split}
\end{equation}
which describes exactly the process (\ref{eq:wealthexchange}) in the
limit $N\to\infty$.  
This equation has the same form as the mean-field version for the 
well-studied 
Maxwell model of inelastically scattering particles
\cite{kra_benna_01,bena_benn_lin_ros_03,ern_bri_02b}.
The main  difference consists in the fact that here the wealth increases,
while in inelastic gas the energy decreases. This seemingly little difference
has, however, deep consequences for the solution of Eq. (\ref{eq:boltzmann}).

Note also that within the framework of Maxwell model the distributions
are expressed in terms of velocities, while our dynamical variables
correspond rather to energies of the particles.

\section{Solution of the kinetic equation}

\subsection{Self-similar solutions}

Note first that the average wealth $\bar{v}=\int v\,P(v)\,\mathrm{d}v$ in
 the process described by the
kinetic equation (\ref{eq:boltzmann}) grows exponentially
\begin{equation}
\bar{v}(\tau)=\bar{v}(0)\,{\rm e}^{\epsilon \tau}
\end{equation}
and therefore Eq. (\ref{eq:boltzmann}) has no stationary solution. However, we 
may look for a quasi-stationary self-similar solution in the form \cite{bob_cer-all,kra_benna_01,bena_benn_lin_ros_03,ern_bri_02b}
\begin{equation}
P(\tau;v)=\frac{1}{\bar{v}(\tau)}\,\Phi(\frac{v}{\bar{v}(\tau)})\; .
\label{eq:scalingform}
\end{equation}
Using the Laplace transform 
$
\hat{\Phi}(x)=\int_0^\infty \Phi(w)\,{\rm e}^{-xw}\,{\rm d}w
$
we can write a non-local differential equation for the scaling
function in the form
\begin{equation}
\epsilon x\hat{\Phi}'(x) +\hat{\Phi}(x)=
\hat{\Phi}((1-\beta+\epsilon)x)\;  
\hat{\Phi}(\beta x)  
\label{eq:nonlocal}
\end{equation}

A hint about possible solutions can be obtained from a special exactly
solvable case $\epsilon=-2\sqrt{\beta}+2\beta$. It can be easily
verified \cite{kra_benna_01} that the function
$\hat{\Phi}_1(x)=(1+\sqrt{2x})\mathrm{e}^{-\sqrt{2x}}$ is a solution of
(\ref{eq:nonlocal}). Inverting the Laplace transform we obtain the
corresponding wealth distribution 
$
{\Phi}_1(w) =$
$ 
\frac {1}{\sqrt {2\pi }}$ ${w}^{-5/2}$ $\exp(-\frac{1}{2\,{w}})
$
which has similar form as obtained in previous studies
\cite{sol_ric_01b,ma_mas_zha_98,bou_mez_00}. 
However, in this case the value of $\epsilon$ is negative,
which contradicts our assumption of wealth increase, while for
$\epsilon>0$ the above idea leading to the function $\hat{\Phi}_1(x)$
does not work. Therefore, we must look for alternative ways. The
leading idea of our approach is that equation (\ref{eq:nonlocal}) is
nearly local for small values of $\epsilon$ and $\beta$. Therefore, we
will expand the factors on the RHS of Eq.  (\ref{eq:nonlocal}) in
Taylor series in $\epsilon$ and $\beta$ and perform the limit
$\epsilon,\beta\to 0$. As the parameters  $\epsilon$ and $\beta$
quantify the amount of wealth increase and exchange in single trade
event, we interpret the latter limit as the limit of continuous
trading. In fact, such limit should involve also a rescaling of time
$\tau$, but because we are interested only in stationary regime, the
explicit time dependence does not enter our considerations.

It should be also stressed that an important feature can be inferred
from the 
observation that the system behaves differently for positive and
negative $\epsilon$. Indeed, it suggests a singularity at the point of
precise conservation of wealth, $\epsilon=0$. 

\begin{figure}[ht]
\includegraphics[scale=0.8]{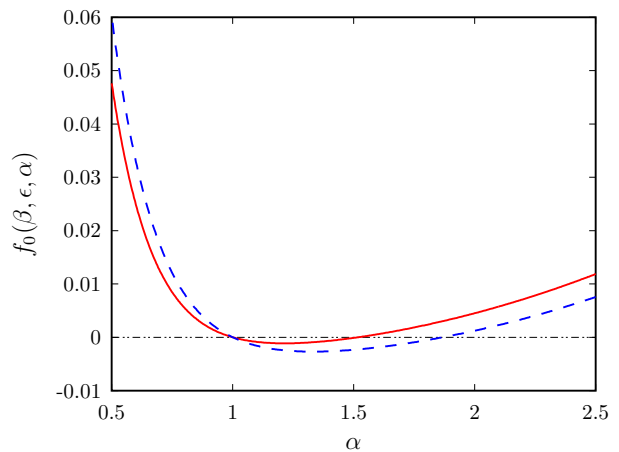}
\caption{%
Solution of the equation $f_0(\beta,\epsilon,\alpha)\equiv
(1+\epsilon-\beta)^\alpha
+\beta^\alpha-1-\epsilon\alpha=0 
$ for $\epsilon=0.1$ and $\beta=0.0025$ (%
full line) and
$\beta=0.004$ 
(dashed line).
}
\label{fig:eqforalpha}
\end{figure}

\subsection{Power-law tails}

\begin{figure}[ht]
\includegraphics[scale=0.8]{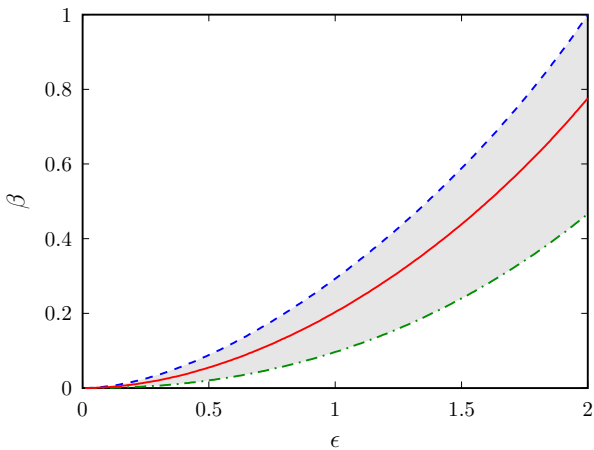}
\caption{%
Solution in the range $\alpha\in(1,2)$ exists within the shaded
region. 
dashed line corresponds to $\alpha=2$,
dash-dotted line corresponds to $\alpha=1$ and
full line to the solution $\alpha=\frac{3}{2}$.
}
\label{fig:phasediagram}
\end{figure}

The main concern in empirical studies of wealth distribution is about
the shape of tails, which assumes power-law form. The behavior of the
distribution $\Phi(w)$ for $w\to\infty$ can be deduced from the
singularity of the Laplace transform $\hat{\Phi}(x)$ at $x\to
0$. Therefore, we assume the following behavior
\cite{kra_benna_01,ern_bri_02b} 
\begin{equation}
\hat{\Phi}(x)=1-x+A\,|x|^\alpha+...\text{ for }x\to 0
\label{eq:smallxsingilarity}
\end{equation}

where $\alpha\in(1,2)$. This type of singularity results in the power-law tail as
${\Phi}(w)\sim w^{-\alpha-1}$ for $w\to\infty$.
Insertion of (\ref{eq:smallxsingilarity}) into (\ref{eq:nonlocal})
leads to a transcendental equation for the exponent $\alpha$
\begin{equation}
(1+\varepsilon-\beta)^\alpha
+\beta^\alpha-1-\varepsilon\alpha=0 
\label{eq:foralpha}
\end{equation}
the solution of which is illustrated in
Fig. \ref{fig:eqforalpha}. Obviously, there is always a trivial
solution $\alpha=1$. The power-law tail is due to another,
non-trivial solution, which falls into the desired interval $(1,2)$
only for certain values of the parameters $\beta$ and $\epsilon$.  We
can see the allowed region in Fig. \ref{fig:phasediagram};
solution in the range $\alpha\in(1,2)$ exists within the shaded
region. We can also see that fixed value of $\alpha$ defines a line in
the $\beta$-$\epsilon$ plane. We can approach
the limit $\epsilon\to 0$, $\beta\to 0$ 
while keeping $\alpha$ constant. This is to be interpreted as
continuous trading, as the amount of wealth exchange and increase in a
single trading step is infinitesimally small. Making this, the non-local terms in
Eq. (\ref{eq:nonlocal}) become local and we can expect to obtain an
ordinary differential equation, soluble by standard methods. 

\subsection{Continuous trading limit}

Indeed, expanding (\ref{eq:foralpha}) we obtain the following 
formula relating $\beta$ and $\epsilon$ for fixed $\alpha$
in the limit of continuous trading
$\beta\to 0$, $\epsilon\to 0$:
\begin{equation}
\beta=\frac{\alpha-1}{2}\,\epsilon^2 +
O(\epsilon^3)+O(\epsilon^{2\alpha})\; .
\label{eq:betaofepsilonlowest}
\end{equation}
The leading correction term to (\ref{eq:betaofepsilonlowest}) depends
on the value of $\alpha$; for $1<\alpha<3/2$ it is of order
$O(\epsilon^{2\alpha})$, for $3/2<\alpha<2$ it is of order
$O(\epsilon^3)$, while in the special point $\alpha=3/2$ we should
include both correction terms, as they are of the same order
$O(\epsilon^3)$. 
Systematic expansion in $\epsilon$ is
developed in Appendix \ref{app:systematicexpansion}.

Taking the same limit with fixed $\alpha$ in
Eq. (\ref{eq:nonlocal}) we obtain, using (\ref{eq:betaofepsilonlowest}),  the following equation
\begin{equation}
-\frac{1}{2}x\hat{\Phi}''(x) +\frac{\alpha-1}{2}\left(\hat{\Phi}'(x) 
+\hat{\Phi}(x)\right)=0  \; .
\label{eq:local}
\end{equation}
Of the two independent solutions of (\ref{eq:local}) only one has 
 correct asymptotics $\hat{\Phi}(x)\to 0$ for
$x\to+\infty$. It can be expressed using modified Bessel function 
\begin{equation}
\hat{\Phi}(x)=C'\,{x}^{\alpha/2\, }
{\it K}_{\alpha}( 2\,\sqrt { \alpha-1}\sqrt {x})
\end{equation}
where the constant $C'$ is fixed by the normalization $\hat{\Phi}(0)=1$.
Inverting the Laplace transform we finally obtain the wealth
distribution
\begin{equation}
\Phi(w)=C\,{w}^{-\alpha-1}\exp(-\frac {\alpha-1}{w})
\label{eq:distribution}
\end{equation}
with $C=(\alpha-1)^\alpha/\Gamma(\alpha)$.

We can see that the distribution obtained exhibits the desired
power-law behavior for large wealth. Moreover, it has a maximum at a
finite value of $w=w_\mathrm{max}\equiv (\alpha-1)/(\alpha+1)$ and
depression for low wealth values. The size of 
the depletion is determined by the exponential term in
(\ref{eq:distribution}), i. e. by 
the same value of $\alpha$ which determines the power in the
power-law. This
corresponds to the idea presented e. g. in Ref. \cite{solomon_99}
stating that it is the value of the lower bound for the allowed wealth
which determines the value of the exponent. Here, however, this result
comes purely formally as a result of the analytic computation. In our
approach it is the interplay between wealth increase (parameter
$\epsilon$) and wealth exchange (parameter $\beta$) that dictates the value of
the exponent $\alpha$. 

\subsection{Corrections for finite trading in one step}

Expanding the equation (\ref{eq:nonlocal}) in powers of
$\epsilon$ and $\beta$ it is possible to include systematic corrections
to equation (\ref{eq:local}) and therefore corrections to wealth
distribution (\ref{eq:distribution}) for finite amount of wealth
increase and exchange in single trading step. Details of the
calculations are given in Appendix \ref{app:systematicexpansion}; here
we only summarize the results.

The expansion (\ref{eq:betaofepsilonlowest}) of the parameter $\beta$ in
powers of $\epsilon$ can be 
continued as
\begin{equation}
\begin{split}
\beta=&\frac{\alpha-1}{2}\,\epsilon^2 +\\ 
&+\frac{1}{\alpha}\left(\frac{\alpha-1}{2}\right)^\alpha\, \epsilon^{2\alpha}
-\frac{(\alpha-1)(2\alpha-1)}{6}\,\epsilon^3+\\
&+O(\epsilon^4)+O(\epsilon^{4\alpha-2})\; .
\label{eq:betaofepsiloncorrected}
\end{split}
\end{equation}
Correspondingly, the wealth distribution, expanded in powers of
$\epsilon$ is
\begin{equation}
\begin{split}
\Phi(w)=&\,\frac{(\alpha-1)^\alpha}{\Gamma(\alpha)}\,w^{-1-\alpha}\exp\left(\frac{1-\alpha}{w}\right)
\times\\&\times
\bigg[1+
\frac{\alpha-1}{3}\,\left(\frac {2\,\alpha}{w}
-\frac{\alpha-1}{w^2}-\nu_{10}\right)\epsilon\,
-\\&-
\frac{2}{\alpha}\,\left(\frac{\alpha-1}{2} \right)
^{\alpha}\,   \left( 
\ln w +\frac {1}{w}-\nu_{01}\right)\epsilon^{2(\alpha-1)}\,
\bigg]
+\\&+
O(\epsilon^4)+O(\epsilon^{4\alpha-2})
\end{split}
\label{eq:distributioncorrected}
\end{equation}
where the constants $\nu_{01}$ and $\nu_{10}$ are given in Appendix 
\ref{app:systematicexpansion}. We show in Fig. \ref{fig:distribution}
the wealth distribution according to (\ref{eq:distributioncorrected})
for $\alpha=1.7$ and several positive values of $\epsilon$, namely for
$\epsilon=0.03$, $0.1$, and $0.3$. We can see that the distribution is
affected mainly at small values of wealth, shifting the maximum
toward smaller $w$ when $\epsilon$ increases. On the contrary, the
tail of the distribution is nearly unaffected, showing universal and
robust power-law behavior.  
\begin{figure}[ht]
\includegraphics[scale=0.8]{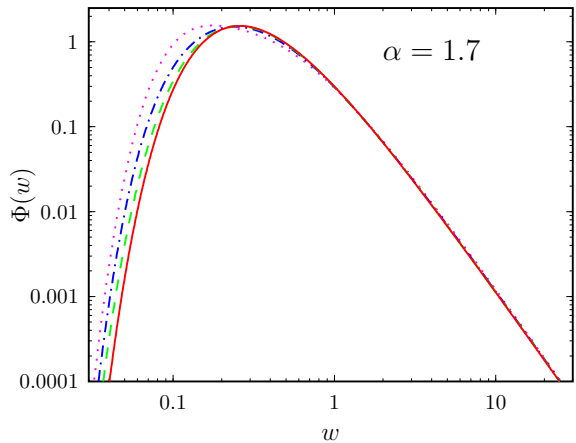}
\caption{%
Wealth distribution according to Eq. (\ref{eq:distributioncorrected})
for $\epsilon\to 0$ (full line), $\epsilon=0.03$ (dashed line), $\epsilon=0.1$
(dash-dotted line), and $\epsilon=0.3$
(dotted line).
}
\label{fig:distribution}
\end{figure}

\begin{figure}[ht]
\includegraphics[scale=0.8]{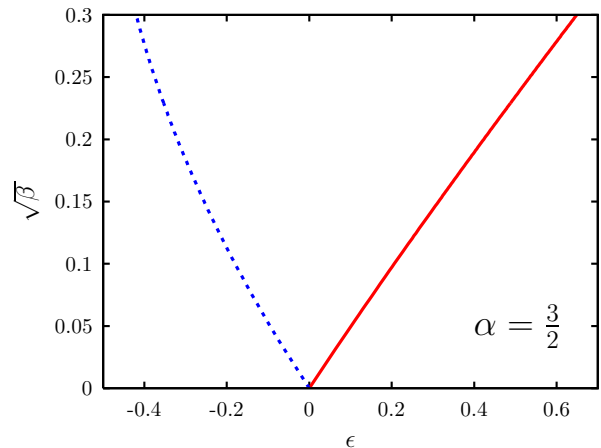}
\caption{%
Solution of equation (\ref{eq:foralpha}) for $\alpha=3/2$ in the ranges
 $\epsilon> 0$ (full line) and $\epsilon<0$ (dashed line). Note the
singularity at $\epsilon=0$ which means that we must skip from one of
the three solutions of  (\ref{eq:foralpha}) to another one.   
}
\label{fig:singularity}
\end{figure}

Let us stress again that the solution known for $\epsilon<0$ cannot be
properly continued to the region of $\epsilon>0$, due to the presence
of singularity at $\epsilon=0$. The singularity can be seeen
e. g. in the behavior of the solution of Eq. (\ref{eq:foralpha}), as
shown in Fig. \ref{fig:singularity}. However, for $\alpha=3/2$ the formula
(\ref{eq:distribution}) describes the solution of  (\ref{eq:nonlocal})
on both limits $\epsilon\to 0^+$ and $\epsilon\to 0^-$. This implies
that  
the singularity is rather weak, because the solution of
Eq. (\ref{eq:nonlocal}) is continuous in $\epsilon$, and only the
derivative with respect of $\epsilon$ has a jump at
$\epsilon=0$. 
One may speculate about the fate of the singularity if we allowed
$\epsilon$ and $\beta$ not fixed parameters but random processes
themselves. Most probably the singularity would vanish but final
answer is left for future work.

\section{Conclusions}

We formulated a model of wealth production and exchange, where agents
randomly interact pairwise. Using the analogy with the mean-field
version of the Maxwell model for inelastic scattering of granular
particles we obtain analytical results for the wealth distribution. 

The dynamics of the model is governed by a kinetic equation for
one-particle distribution function. We look  for self-similar
scaling solutions, corresponding to redefining the unit of wealth
after each wealth increase. The form of these solutions is given by a
non-local differential equation, exactly soluble only in the
practically irrelevant case of net wealth decrease. Therefore we
turned to approximation schemes.

First, we looked at the behavior for large wealth.
The tail of the wealth distribution has a power-law form, and its
exponent $\alpha$ is determined by the interplay between the intensity of 
the wealth exchange and the amount of wealth produced. The form line in the
$\beta$-$\epsilon$ plane with
fixed $\alpha$ is found, depending quadratically on $\epsilon$ for
$\epsilon\to 0$. The physically allowed values $\alpha\in(1,2)$
determine a horn-shaped region in the $\beta$-$\epsilon$ plane.

The second approximation consisted in taking the limit of continuous
trading, meaning small wealth
production and small exchange within a single trading operation, 
while keeping the exponent $\alpha$
constant. Here we obtained closed formula for the entire wealth
distribution, which has power-law tail as expected and a maximum at
certain (low) wealth value. The form of the wealth distribution
corresponds to those found in previous studies 
\cite{sol_ric_01b,ma_mas_zha_98,bou_mez_00}. It is interesting to note
that this general form has one-to-one correspondence between the
position $w_\mathrm{max}$ of the maximum of the distribution and the
value of the 
exponent. There are few agents having wealth below $w_\mathrm{max}$. 
 This suggests that the intuition formalized e. g. in
\cite{sol_ric_01b,solomon_99}, that the exponent is ``tuned'' by the
low-wealth behavior  of the distribution, may be in work quite
generally. Here, the free parameters are apparently the wealth
production and exchange, but in reality these parameters may be
themselves tuned by a mechanism which fixes the position of the
maximum of the wealth distribution, i. e. the lowest wealth compatible
with survival. 

However, there is still open question of the specific values of the
exponent, which are quite robust in different societies. It seems,
also on the basis of our results, that it cannot be explained by the
bare mechanism of economic exchange and some other ingredient,
possibly of sociological origin, is required.
\begin{acknowledgments}
I wish to thank  Paul Krapivsky and Eli Ben-Naim 
for stimulating comments
and discussions.
This work was supported by the project No. 202/01/1091
of the Grant Agency  of the Czech Republic. 
\end{acknowledgments}
\vspace*{2mm}
\appendix

\section{ Systematic expansion for small $\epsilon$ and $\beta$.}
\label{app:systematicexpansion}

Let us start with the special value
 $\alpha=3/2$. Here, the equation (\ref{eq:foralpha}) has an explicit
 solution in the form   
\begin{widetext}
\begin{equation}
\epsilon=\frac{1}{8}\;{\frac {-3\,\sqrt {\beta}+17\,\beta-29\,{\beta}^{3/2}+15\,{\beta}^{2}+4\,{\beta}^{5
/2}-4\,{\beta}^{3}+
\sqrt {3}\sqrt {\left( 3-2\,\sqrt {\beta} \right) \beta\,
 \left( 2\,\sqrt {\beta}+1 \right) ^{3} \left( \sqrt {\beta}
-1 \right) ^{6}}}{\sqrt {\beta}-3\,\beta+3\,{\beta}^{3/
2}-{\beta}^{2}}}\; .
\label{eq:epsofbetathreehalfs}
\end{equation}
\end{widetext}
However, the non-local
differential equation  (\ref{eq:nonlocal}) still does not yield 
explicit solution.
Inverting the expression (\ref{eq:epsofbetathreehalfs}) we get the
following series expansion
\begin{equation}
\beta={\frac {1}{4}}{\epsilon}^{2}-{\frac {1}{12}}{\epsilon}^{3}+
{\frac {1}{16}}{\epsilon}^{4}-{\frac {7}{144}}{\epsilon}^{5}
+{\frac {113}{2592}}{\epsilon}^{6}+O \left( {\epsilon}^{7} \right)\, .  
\end{equation}

For general value of $\alpha$ the variable $\beta$ is expressed as a
series in two small parameters $\epsilon$ and
$\eta=\epsilon^{2(\alpha-1)}$, which coincide only if
$\alpha=3/2$. Therefore, we can write
\begin{equation}
\beta=\epsilon^2\,\sum_{m,n=0}^\infty\beta_{mn}\epsilon^{m+2(\alpha-1)n}
\label{eq:expansionforbeta}
\end{equation}
and the various terms take variable precedence in the order of
smallness when $\epsilon\to 0$, depending on the value of
$\alpha$. For the first several coefficients we have
\begin{eqnarray}
\beta_{00}&=&\frac{\alpha-1}{2}\\ 
\beta_{10}&=&-\frac{(\alpha-1)(2\alpha-1)}{6}\\
\beta_{01}&=&\frac{1}{\alpha}\left(\frac{\alpha-1}{2}\right)^\alpha\; .
\end{eqnarray}
Starting from the expansion (\ref{eq:expansionforbeta}) we can convert
the first order non-local differential equation (\ref{eq:nonlocal}) for
$\hat{\Phi}(x)$ into infinite-order local differential equation for
$\Phi(w)$. The price to pay for it is that the coefficients in the
latter equation contain the moments $\mu_k=\int\Phi(w)w^k\mathrm{d}w$ 
of the solution itself. Indeed, we
can write
\begin{equation}
\hat{\Phi}((1-\beta+\epsilon)\,x)=\lim_{y\to x}
\exp\left(
(\epsilon-\beta)\,x\frac{\mathrm{d}}{\mathrm{d}y}
\right)\,
\hat{\Phi}(y)
\end{equation}
\begin{equation}
\hat{\Phi}(\beta x)=\lim_{y\to 0}\exp\left(
\beta\,x\frac{\mathrm{d}}{\mathrm{d}y}
\right)\,
\hat{\Phi}(y)\;.
\end{equation}
Therefore, we obtain a linear combination of terms of the following
form
\begin{equation}
x^{m+n}\,\frac{\mathrm{d}^m\hat{\Phi}(x)}{\mathrm{d}x^m}\,\frac{\mathrm{d}^n\hat{\Phi}(0)}{\mathrm{d}x^n}
\end{equation}
which, after inverse Laplace transform, give rise to terms
\begin{equation}
(-1)^{m+n}\,\mu_n\,\frac{\mathrm{d}^{m+n}}{\mathrm{d}w^{m+n}}\,
\left[
w^m\,\Phi(w)
\right]\; .
\end{equation}

However,
 the first two moments are fixed by definition. Indeed, the
normalization of the probability distribution fixes the zeroth moment
and the fixed average wealth, imposed by the scaling condition
(\ref{eq:scalingform}) fixes the first moment, so that
$\mu_0=\mu_1=1$. 
This consideration
leads to the equations for lowest correction to the solution
(\ref{eq:distribution}), which are free of unknown higher moments.

Generally, the solution can be then expressed in the form of the
series in powers of $\epsilon$ and $\epsilon^{2(\alpha-1)}$
\begin{equation}
\Phi(w)=\Phi_0(w)\sum_{m,n=0}^\infty\phi_{mn}(w)\,\epsilon^{m+2(\alpha-1)n}\; .
\end{equation}
We assume $\phi_{00}(w)=1$. The normalization must be independent of 
 $\epsilon$, which can be written as
\begin{equation}
\int_0^\infty\Phi_0(w)\phi_{mn}(w)\mathrm{d}w =\delta_{m0}\delta_{n0}\; .
\label{eq:normalizationforalleps}
\end{equation}

Therefore, the lowest term obeys the
equation
\begin{equation}
\frac{w^2}{2}\Phi_0'(w)+\left(\frac{\alpha+1}{2}w-\frac{\alpha-1}{2}\right)
\Phi_0(w)=0
\end{equation}
which has the following solution satisfying the normalization 
(\ref{eq:normalizationforalleps})
\begin{equation}
\Phi_0(w)=\frac{(\alpha-1)^\alpha}{\Gamma(\alpha)}\,w^{-1-\alpha}\exp\left(\frac{1-\alpha}{w}\right)\; .
\end{equation}
Indeed, it coincides with the result of (\ref{eq:distribution}).

The next two terms satisfy the following equations
\begin{eqnarray}
&&\frac{w^2}{2}\phi_{10}'(w)=\frac{\alpha-1}{3}\,
\left(\alpha-
 \frac{\alpha-1}{w}
\right)
\\
&&\frac{w^2}{2}\phi_{01}'(w)=-\frac{1}{\alpha}\,\left(\frac{\alpha-1}{2} \right)
^{\alpha}\,   \left( w-1\right)
\end{eqnarray}
which can be easily solved. We obtain
\begin{eqnarray}
\phi_{10}(w)&=&-\frac{\alpha-1}{3}\,\left(\frac {2\,\alpha}{w}
-\frac{\alpha-1}{w^2}-\nu_{10} \right) 
\\
\phi_{01}(w)&=&-\frac{2}{\alpha}\left(\frac{\alpha-1}{2} \right)
^{\alpha}   \left( 
\ln w +\frac {1}{w}
-\nu_{01}
\right)
\end{eqnarray}
and the constants $\nu_{01}, \nu_{10}$ are fixed by the normalization
condition  
(\ref{eq:normalizationforalleps}). We
 find explicitly 
\begin{eqnarray}
\nu_{10}&=&\alpha\\
\nu_{01}&=& 
\ln  \left( \alpha-1 \right) - 
\Psi \left( \alpha \right) + \frac{\alpha}{{\alpha}-1 } 
\end{eqnarray}
where $\Psi(x)=\Gamma'(x)/\Gamma(x)$ is the logarithmic derivative of
the gamma function.

{}

\begin{thebibliography}{99}
%
\bibitem{pareto_1897}
V. Pareto,
 {\it Cours d'economie politique}, 
(Lausanne, F. Rouge, 1897).

\bibitem{lev_sol_97}
M. Levy and S. Solomon,
 Physica A
{\bf 242}, 90
(1997).

\bibitem{dra_yak_01a}
A. Dr\u agulescu and V. M. Yakovenko,
 Physica A
{\bf 299}, 213
(2001).

\bibitem{ree_hug_02}
W. J. Reed and B. D. Hughes,
 Phys. Rev. E
{\bf 66}, 067103
(2002).

\bibitem{aoy_sou_fuj_03}
H. Aoyama, W. Souma, and Y. Fujiwara,
 Physica A
{\bf 324}, 352
(2003).

\bibitem{abulmagd_02}
A. Y. Abul-Magd,
 Phys. Rev. E
{\bf 66}, 057104
(2002).

\bibitem{lev_sol_96}
M. Levy and S. Solomon,
 Int. J. Mod. Phys. C
{\bf 7}, 595
(1996).

\bibitem{lev_sol_96a}
M. Levy and S. Solomon,
 Int. J. Mod. Phys. C
{\bf 7}, 65
(1996).

\bibitem{bi_ma_le_so_98}
O. Biham, O. Malcai, M. Levy, and S. Solomon,
 Phys. Rev. E
{\bf 58}, 1352
(1998).

\bibitem{solomon_98}
S. Solomon,
 in: {\it Decision technologies for Computational Finance}, ed. A.-P. Refenes, A. N. Burgess, and J. E. Moody
(Kluwer Academic Publishers, 1998).

\bibitem{solomon_99}
S. Solomon,
 {\rm in: {\it Application of Simulation to Social Sciences}, ed. G. Ballot and G. Weisbuch} 
(Hermes Science Publications, 2000).

\bibitem{hua_sol_00}
Z.-F. Huang and S. Solomon,
 Eur. Phys. J. B
{\bf 20}, 601
(2001).

\bibitem{sol_ric_01b}
S. Solomon and P. Richmond,
 Physica A
{\bf 299}, 188
(2001).

\bibitem{bla_sol_00}
A. Blank and S. Solomon,
 Physica A
{\bf 287}, 279
(2000).

\bibitem{sol_lev_00}
S. Solomon and M. Levy,
 cond-mat/0005416.

\bibitem{hua_sol_01}
Z.-F. Huang and S. Solomon,
 Physica A
{\bf 294}, 503
(2001).

\bibitem{so_co_97}
D.~Sornette and R.~Cont,
 J. Phys I France
{\bf 7}, 431
(1997).

\bibitem{sornette_97a}
D.~Sornette,
 Physica A
{\bf 250}, 295
(1998).

\bibitem{sornette_98c}
D. Sornette,
 Phys. Rev. E
{\bf 57}, 4811
(1998).

\bibitem{ta_sa_ta_97}
H. Takayasu, A.-H. Sato, and M. Takayasu,
 Phys. Rev. Lett.
{\bf 79}, 966
(1997).

\bibitem{ma_mas_zha_98}
M. Marsili, S. Maslov, and Y.-C. Zhang,
 Physica A
{\bf 253}, 403
(1998).

\bibitem{bou_mez_00}
J.-P. Bouchaud and M. M\'ezard,
 Physica A
{\bf 282}, 536
(2000).

\bibitem{bur_joh_jur_kam_nov_pa_zah_01}
Z. Burda, D. Johnston, J. Jurkiewicz, M. Kami\'nski, M. A. Nowak, G. Papp, and I. Zahed,
 cond-mat/0101068.

\bibitem{souma_00}
W. Souma,
 cond-mat/0011373.

\bibitem{aoy_nag_oka_sou_tak_tak_00}
H. Aoyama, Y. Nagahara, M. P. Okazaki, W. Souma, H. Takayasu, M. Takayasu,
 cond-mat/0006038.

\bibitem{souma_02}
W. Souma,
 cond-mat/0202388.

\bibitem{sou_fuj_aoy_01}
W. Souma, Y. Fujiwara, and H. Aoyama,
 cond-mat/0108482.

\bibitem{fuj_sou_aoy_kai_aok_02}
Y. Fujiwara, W. Souma, H. Aoyama, T. Kaizoji, and M. Aoki,
 cond-mat/0208398.

\bibitem{cha_cha_00}
A. Chakraborti and B. K. Chakrabarti,
 Eur. Phys. J. B
{\bf 17}, 167
(2000).

\bibitem{cha_cha_03}
B. K. Chakrabarti and A. Chatterjee,
in: {\it Applications
of Econophysics}, Conference proceedings of Second 
Nikkei Symposium on Econophysics, Tokyo, Japan, 2002 
(Springer-Verlag, Tokyo, 2003)  pp. 280-285;
 cond-mat/0302147.

\bibitem{cha_cha_man-all}
A. Chatterjee, B. K. Chakrabarti, and S. S. Manna,
 cond-mat/0301289, to be published in Physica A;
Phys. Scripta T106, 36 (2003); cond-mat/0311227.

\bibitem{das_yar_03}
A. Das and S. Yarlagadda,
 cond-mat/0304685.

\bibitem{pia_igl_abr_veg_01}
S. Pianegonda, J. R. Iglesias, G. Abramson, and J. L. Vega,
 Physica A
{\bf 322}, 667
(2003).

\bibitem{igl_gon_pia_veg_abr_03}
J. R. Iglesias, S. Gon\c calves, S. Pianegonda, J. L. Vega, and G. Abramson,
 Physica A
{\bf 327}, 12
(2003).

\bibitem{pia_igl_03}
S. Pianegonda and J. R. Iglesias,
 cond-mat/0311113.

\bibitem{igl_gon_abr_veg_03}
J. R. Iglesias, S. Gon\c calves, G. Abramson, and  J. L. Vega,
 cond-mat/0311127.

\bibitem{ana_ish_suz_tom_03}
M. Anazawa, A. Ishikawa, T. Suzuki, and M. Tomoyose,
 cond-mat/0307116.

\bibitem{miz_kat_tak_tak_03}
T. Mizuno, M. Katori, H. Takayasu, and M. Takayasu,
 cond-mat/0308365.

\bibitem{miz_tak_tak_03}
T. Mizuno, M. Takayasu, and H. Takayasu,
 cond-mat/0307270.

\bibitem{fuj_dig_aoy_gal_sou_03}
Y. Fujiwara, C. Di Guilmi, H. Aoyama, M. Gallegati, and W. Souma,
 cond-mat/0310061.

\bibitem{dra_yak-all}
A. Dr\u agulescu and V. M. Yakovenko,
 Eur. Phys. J. B
{\bf 17}, 723
(2000);
 Eur. Phys. J. B
{\bf 20}, 585
(2001);
 in: {\it Modeling of Complex Systems: Seventh Granada Lectures}, AIP Conference Proceedings 661
{\bf }, 180
(New York, 2003).

\bibitem{yakovenko_03}
V. M. Yakovenko,
 cond-mat/0302270.

\bibitem{is_kra_re_98}
S. Ispolatov, P. L. Krapivsky, and S. Redner,
 Eur. Phys. J. B
{\bf 2}, 267
(1998).

\bibitem{sca_pic_wes_02}
N. Scafetta, S. Picozzi, and B. J. West,
 cond-mat/0209373.

\bibitem{sca_wes_03}
N. Scafetta and B. J. West,
 cond-mat/0306579.

\bibitem{gli_ign_02}
M. Gligor and M. Ignat,
 Eur. Phys. J. B
{\bf 30}, 125
(2002).

\bibitem{sinha_03}
S. Sinha,
 cond-mat/03043224.

\bibitem{ja_na_be_96a}
H. M. Jaeger, S. R. Nagel, and R. P. Behringer,
 Rev. Mod. Phys.
{\bf 68}, 1259
(1996).

\bibitem{bob_car_gam_00}
A. V. Bobylev, J. A. Carillo, and I. M. Gamba,
 J. Stat. Phys.
{\bf 98}, 743
(2000).

\bibitem{bob_cer-all}
A. V. Bobylev and C. Cercignani,
 J. Stat. Phys.
{\bf 106}, 547
(2002);
 J. Stat. Phys.
{\bf 110}, 333
(2003).

\bibitem{bal_mar_pug_01a}
A. Baldassarri, U. Marini Bettolo Marconi, and A. Puglisi,
 Europhys. Lett.
{\bf 58}, 14
(2002).

\bibitem{isp_kra_00}
I. Ispolatov and P. L. Krapivsky,
 Phys. Rev. E
{\bf 61}, R2163
(2000).

\bibitem{benn_kra_00}
E. Ben-Naim and P. L. Krapivsky,
 Phys. Rev. E
{\bf 61}, R5
(2000).

\bibitem{kra_benna_01}
P. L. Krapivsky and E. Ben-Naim,
 J. Phys. A: Math. Gen.
{\bf 35}, L147
(2002).

\bibitem{benn_kra_02a}
E. Ben-Naim and P. L. Krapivsky,
 Phys. Rev. E
{\bf 66}, 011309
(2002).

\bibitem{bena_benn_lin_ros_03}
D. ben-Avraham, E. Ben-Naim, K. Lindenberg, and A. Rosas,
Phys. Rev. E 68, 050103 (2003).

\bibitem{ern_bri_02b}
M. H. Ernst and R. Brito,
 Europhys. Lett.
{\bf 58}, 182
(2002).

\bibitem{ern_bri-everything}
M. H. Ernst and R. Brito,
 cond-mat/0111093;
 Phys. Rev. E
{\bf 65}, 040301
(2002);
 J. Stat. Phys. 
{\bf 109}, 407 
(2002);
 Europhys. Lett.
{\bf 58}, 182
(2002);
 cond-mat/0304608.



%
%
%
%

\bibitem{benn_kra_03}
E. Ben-Naim and P. L. Krapivsky,
 cond-mat/0301238.

\bibitem{ant_dro_lip_02}
T. Antal, M. Droz, and A. Lipowski,
 Phys. Rev. E
{\bf 66}, 062301
(2002).

\bibitem{barkai_03}
E. Barkai, 
Phys. Rev. E 68, 055104 (2003).


\bibitem{bal_mar_pug_01}
A. Baldassarri, U. Marini Bettolo Marconi, and A. Puglisi,
 cond-mat/0105299.

\bibitem{bal_mar_pug_vul_01}
A. Baldassarri, U. Marini Bettolo Marconi, A. Puglisi, and A. Vulpiani,
 Phys. Rev. E
{\bf 64}, 011301
(2001).

\bibitem{wa_stro_98}
D. J. Watts and S. H. Strogatz,
 Nature
{\bf 393}, 440
(1998).

\bibitem{bar_alb_99}
A.-L. Barab\'asi and R. Albert,
 Science
{\bf 286}, 509
(1999).

\bibitem{dim_ast_hyd_03}
T. Di Matteo, T. Aste, and S. T. Hyde,
 cond-mat/0310544.

%
\end{thebibliography}
\end{document}